# AN INNOVATIVE SQA SERVICE MATURITY MODEL USING CMMI AND ITIL

Shankar Gurumoorthy

Senior Technical Project Leader, Bangalore, India

shankar.gtech@gmail.com

**ABSTRACT**

*This Journal details a maturity model for SQA services which has been developed during QMS implementation in the IT division of a large multinational organization.*

*The scope of the engagement was to establish a standard set of processes based on CMMI® and ITIL® Framework across four business verticals scattered in Europe, United States and Asia. The services of Software Quality Analyst (SQA) from different vendors were leveraged to facilitate implementation of processes which was referred to as the Quality Management System (QMS).*

*To co-ordinate and support QMS implementation, a Software Quality Assurance Group (SQAG) was established at the organizational level. The role of SQAs was to facilitate the deployment of QMS by mentoring and guiding practitioners through the software development and maintenance lifecycle, identify deviations and addressing the same, based on CMMI and developed to eventually conduct periodic process compliance checks to verify process implementation at pre-defined intervals, collate feedback from practitioners on challenges faced in implementation, identify process improvement opportunities and implement the same.*

*Considering the large number of applications, the business verticals proposed that process implementation should be owned and managed by practitioners themselves so that the mass deployment of QMS can be achieved at a faster rate with the same SQA capacity. Considering that the process framework was developed by vendor and subject matter experts, it was important for practitioners to be trained and handheld on processes prior to being empowered with the accountability for adherence to QMS. The risk of reduced independent insight of management on process implementation was highlighted. Additionally, risk of implementing QMS without adequate training and experience of practitioners was also communicated.*

*This called for a need to devise an innovative implementation solution before moving to a process implementation model which proposed Project Managers implementing processes themself. While there are process models and frameworks available in the market for establishing processes in an organization, there is no model that elaborates activities to be performed by the SQA for effective implementation of processes. SQA service maturity model was proposed as a solution based on CMMI® and developed to eventually proceed towards a 'Process Implementation Model proposing Project Managers implementing processes themself'.*

*This SQA service maturity model comprises of five levels. At maturity level 1, there is no SQA organization in place and roles & responsibilities are not defined Also, there are no quality standards/models followed for delivering SQA services. At maturity level 2, SQAG organization formed at the organizational level and services delivered. i.e., Internal Process Consulting services delivered by SQAs. Also, SQAs to facilitate the deployment of QMS by mentoring and guiding practitioners. At this level, basic Quality Assurance processes are implemented. i.e., Software Quality Assurance Lifecycle is implemented. From the SQA technology perspective, SQA document archival Repository created and basic set of SQA activities related tools implemented. i.e., Dashboard for Metrics reporting. At maturty level 3, Management system defined at the organizational level updated to include processes on delivering SQA services. At maturity level 4, SQA service related processes are expected to be automated.*

*Additionally metrics baselines for SQA services are also required to be derived based on historical data. At maturity level 5, Project Managers are expected to be empowered to play role of SQA as well and take accountability of process implementation. The type of process implementation at this level would be referred to as the 'Process Implementation Model managed by practitioners' which would ensure institutionalization of processes at a faster rate. It was explained to the customer that the benefit of the aforesaid model in terms of mass deployment of processes can only be derived when the organization reaches maturity level 5 of SQA service maturity model.*

*SQA Service Maturity Model is a Software Quality Assurance implementation framework that enables organisations to increase Efficiencies in Software Quality Assurance, reduce the Cost of Defects and ultimately Increasing Return on Investment in IT.*

*To optimize and improve the efficiency of SQA services, we have to look at 3 basic pillars - People, Process and Technology. SQA services Maturity Model is a tried and tested, simple and effective methodology that allows you to achieve quick wins in quality assurance by enabling you to deliver quality without compromise. It fits into an agile or traditional development framework.*

*Utilization of the SQA services Maturity Model will enable the organisation to achieve the following:*
*PEOPLE: SQA team with practitioners who are skilled & competent and meet performance targets*
*TECHNOLOGY & DISCIPLINES: End-to-end SQA services are performed*
*PROCESS: SQA practices are optimized improving the service capability. Comprehensive SQA dashboards are delivered to management.*

*The organization is satisfied as SQA effort is reduced and Practitioner awareness on processes and deep product quality aspects are improvised. The ultimate objective is to improve the efficiency of SQA services and the organization cannot achieve this without focusing on People, Processes and Technology.*

*It is recommended that the SQA service maturity model be used as a collection of best practices by organizations setting up a Software Quality Assurance Group (SQAG) to drive process implementation in a phased manner. The defined model will provide a framework to empower organizations to choose the appropriate level of SQA services for implementation and institutionalization of defined processes. While this model can serve as a reference model to begin with, practices at maturity level 4 and maturity level 5 would also help organizations to reduce cost of SQA activities.*

***Objective:*** *The objective of this Journal is to explain the benefits of SQA services maturity model which was used as a reference framework to reiterate to the customer organization that mass deployment of QMS processes can only be achieved after a certain level of process maturity is attained. This model also describes the preliminary set of SQA activities that are involved in establishing the SQA group in an organization before moving to a 'Practitioner Managed Process Implementation'.*

## 1. THE PROCESS IMPROVEMENT INITIATIVE:

This Journal explains a maturity model for SQA services which has been developed during a process consulting engagement with a large multinational organization with IT units located at several locations in United States, Europe and Asia.

The engagement commenced with process definition based on CMMI and ITIL framework and subsequent implementation of the defined processes in a phased manner. Processes that were defined were aligned to the strategic business goals & objectives, organizational standards and policies. Software Quality Assurance Group (SQAG) was established for implementing processes that were defined by the Software Engineering Process Group (SEPG). It was deliberated that SQA group will have a reporting channel to the Senior Management that is independent of practitioners.

Software Quality Resource Capacity Planning was required to be done depending on the number of ongoing projects and size of the projects in each vertical. Based on the verticals and a high level estimation of the effort required to support process implementation, SQAs were assigned to verticals. This process also involved the project/program management to agree with the SQA Group Lead on the tasks that SQA will perform.

SQAG Plan was prepared with detailed responsibilities of SQA activities such as:

- SQA's participation in establishing the procedures specific to the project
- Compliance checks to be done by the SQA using the processes defined by the software engineering process group as basis
- Procedures for documenting and tracking non-compliances and determine corrective/preventive actions
- Method and frequency of providing the process implementation status of projects
- Method and frequency of providing feedback to the software engineering process group for process improvement
- Deliverables that are required to be produced by SQAs

SQAs were trained on defined processes to help them adequately perform the SQA function and also train practitioners on QMS.

SQAs facilitated the deployment of processes at onsite locations by:

- Mentoring and guiding practitioners throughout the software development and maintenance lifecycle
- Identifying deviations and address the same as per the documented procedure
- Conducting periodic process compliance checks to verify implementation of processes
- Collating feedback from practitioners on difficulties faced in implementation and effectiveness of processes towards improvement of existing processes
- Implementing improvement opportunities identified
- Providing independent insight on process implementation to the management at an appropriate level of abstraction and in a timely manner including project status, problem areas, and risks
- Providing additional support to the project manager during senior management reviews to emphasize areas that require management attention.

Assistance was provided in initiating the deployment by participating in training the practitioners on defined processes. The phased deployment approach resulted in process deployment across an increasing number of applications.

Initially, the SQA facilitation was done Onsite by SQA's from another vendor organization. At the end of every year, a target count of applications adopting Quality Management System (QMS) was discussed and agreed upon between the Quality Management Group and Senior Management of IT units. Keeping in mind the increasing number of applications adopting QMS, it became difficult to do process facilitation with the existing strength of SQAs at onsite. This is when the proposal to set up a team of SQAs at offshore was shared with the customer and accepted by them.

## 2. ONSITE-OFFSHORE MODEL:

As part of setting up this model, the list of SQA activities which could be done from offshore were identified under the following categories: Training, Facilitation, Monitoring & Control, Status Reporting, Process Compliance Reviews and Deliverable Reviews. This was discussed, agreed upon with the customer and piloted and subsequently implemented. This model was accepted by the customer organization as it was cost-effective and enabled them to increase the number of projects following QMS by allowing for onsite SQAs to facilitate additional projects.

## 3. Mass Deployment of processes:

One of the verticals in the customer organization wanted to expedite QMS deployment. They asked for a significant increase in the number of projects following the QMS without any increase in the size of the SQA pool. This would not have been possible to implement using the existing process maturity and SQA strength. It was apparent that there was a need to devise a solution that called for lesser SQA interaction with practitioners. Also, Project Managers would have to take responsibility for ensuring that requirements of the QMS were adhered to. This approach ran the risk that projects may deviate significantly from the defined process, which could result the organizational objectives not being met. At this point in time, keeping in mind the challenges with respect to the existing process maturity of the organization and the limited SQA capacity, SQA Service Maturity Model was proposed by us to convey to the management that Project Managers should be held accountable for QMS deployment only when the organization attains a certain level of maturity.

## 4. Why SQA Service Maturity Model:

Since there were no existing models or frameworks available in the market that elaborate the activities that will need to be performed by the SQA for effective implementation of process in a staged manner, SQA Service Maturity model was defined. The objective of this model is to provide a framework that will empower organizations to choose the level of SQA services for implementation and institutionalization of defined processes.

## 5. Principle of SQA Service Maturity Model:

While maturity levels apply to an organization's process improvement achievement, the same can be applied to the services of an SQA. Maturity Levels are used to describe an evolutionary path recommended for organizations that wants to improve the process maturity by leveraging SQA services. The SQA Maturity level of an organization provides a way to predict an organizations' performance with respect to SQA services. An SQA Service Maturity Level is a defined evolutionary plateau for organizational process improvement with respect to SQA services. Also, our experience in implementing this model has shown that each level matures an important subset of the SQA services preparing it to move to the next maturity level.

SQA Service Maturity Model is a Software Quality Assurance implementation framework that enables organisations to increase Efficiencies in Software Quality Assurance, reduce the Cost of process compliance check effort and ultimately Increasing Return on Investment in IT.

To optimize and improve the efficiency of SQA services, we have to look at 3 basic pillars - People, Process and Technology. SQA services Maturity Model is a tried and tested, simple and effective methodology that allows you to achieve quick wins in quality assurance by enabling you to deliver quality without compromise. It fits into an agile or traditional development framework.

Utilization of the SQA services Maturity Model will enable the organisation to achieve the following:

PEOPLE: SQA team with practitioners who are skilled & competent and meet performance targets

TECHNOLOGY & DISCIPLINES: End-to-end SQA services are delivered.

PROCESS: SQA practices are optimized improving the service capability. Comprehensive QA dashboards are delivered to management.

The organization is satisfied as SQA effort is reduced and Practitioner awareness on processes and deep product quality aspects are improvised.

The ultimate objective is to improve the efficiency of SQA services and the organization cannot achieve this without focusing on People, Processes and Technology

As per this model, there are five maturity levels, each a layer in the foundation for ongoing process improvement. These have been designated by numbers 1 through 5:

1. Initial
2. Managed
3. Defined
4. Quantitatively Managed and services automated
5. Optimizing

## 6. MATURITY LEVELS:

### Maturity Level 5 - Optimizing

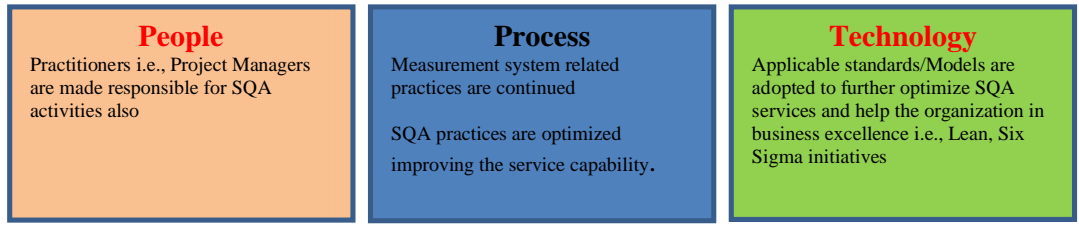

| People | Process | Technology |
|---|---|---|
| Practitioners i.e., Project Managers are made responsible for SQA activities also | Measurement system related practices are continued. SQA practices are optimized improving the service capability. | Applicable standards/Models are adopted to further optimize SQA services and help the organization in business excellence i.e., Lean, Six Sigma initiatives |

### Maturity Level 4 – Quantitatively Managed and services automated

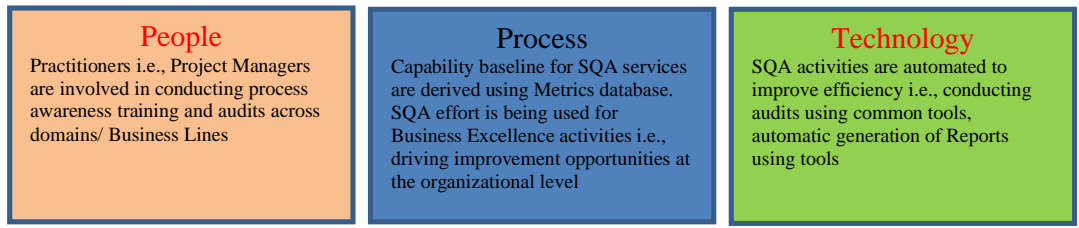

| People | Process | Technology |
|---|---|---|
| Practitioners i.e., Project Managers are involved in conducting process awareness training and audits across domains/ Business Lines | Capability baseline for SQA services are derived using Metrics database. SQA effort is being used for Business Excellence activities i.e., driving improvement opportunities at the organizational level | SQA activities are automated to improve efficiency i.e., conducting audits using common tools, automatic generation of Reports using tools |

### Maturity Level 3 - Defined

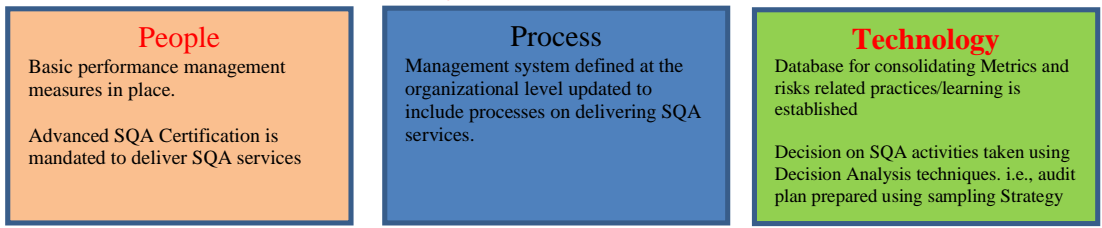

| People | Process | Technology |
|---|---|---|
| Basic performance management measures in place. Advanced SQA Certification is mandated to deliver SQA services | Management system defined at the organizational level updated to include processes on delivering SQA services. | Database for consolidating Metrics and risks related practices/learning is established. Decision on SQA activities taken using Decision Analysis techniques. i.e., audit plan prepared using sampling Strategy |

### Maturity Level 2 - Managed

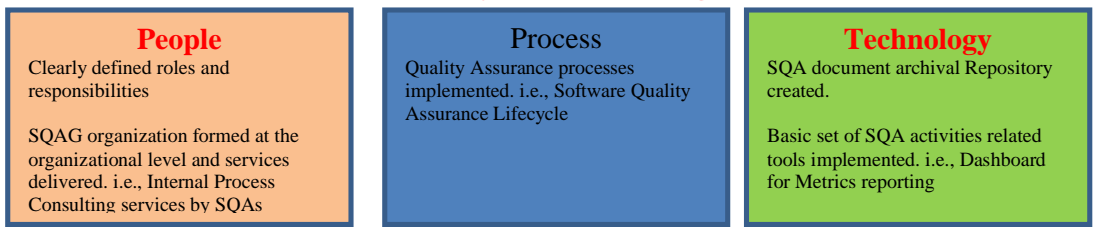

| People | Process | Technology |
|---|---|---|
| Clearly defined roles and responsibilities. SQAG organization formed at the organizational level and services delivered. i.e., Internal Process Consulting services by SQAs | Quality Assurance processes implemented. i.e., Software Quality Assurance Lifecycle | SQA document archival Repository created. Basic set of SQA activities related tools implemented. i.e., Dashboard for Metrics reporting |

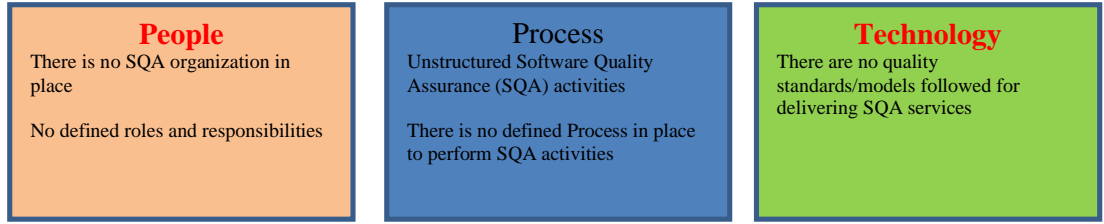

## 6.1 Maturity Level 1 - Initial

At maturity level 1, no processes exist at the organizational level. Even if they exist, they are executed in an adhoc manner and are not planned and tracked. SQA services are thereby not leveraged at this level.

## 6.2 Maturity Level 2

At level 2, requirements from Verticals/Departments for SQA Services are collated and provided to Software Quality Assurance Group. This could include specific requirements from the vertical or departments, if any. SQA service requests are collated and maintained by SQAG. Based on the size (in Person Months), complexity and duration of the project, effort estimation is arrived at by SQAG. Accordingly, SQAs are assigned to the vertical/department. In situations where the staffing of SQAs cannot be done from the organizations' internal resource pool, services of an SQA can be acquired from vendor, in accordance with sourcing policies of the organization. SQAG prepares a plan referred as "SQAG Plan" which exists at an organization level. SQAG Plan is a catalogue of all possible services provided by the SQA.

When the SQA starts engaging with projects, SQA Plan is prepared by each SQA. This plan provides information on on-boarding of SQAs, list of activities to be done by SQAs driven by milestones and any specific training which SQA needs to attend. Training requirement could pertain to process training/tool training requirement. At this level, potential risk with respect to performance of SQA activities need to be identified and tracked. Risks could be owing to any of the following – lack of interest in process implementation amongst practitioners, project teams' bandwidth issues, language constraints, aggressive timelines, budget constraints etc.

Project monitoring involves tracking of the actual effort against planned in the SQA plan. In case of deviations in SQA plan, corrective actions need to be identified and implemented. SQA willbe involved in evaluation of project/process deliverables from content and completeness perspective and in conducting process compliance checks to gauge the compliance to the organizations' defined set of process and reporting the findings.

This level emphasizes the need to adopt Configuration Management procedures for deliverables produced and maintained by SQAs. This includes SQA plan, compliance check reports/dashboard, gap analysis report, project metric analysis report etc. This could be maintained in a central repository by providing necessary access rights

Measurement and analysis activities with respect to SQA services should be aligned to Key Performance Indicators required for management reporting. Additionally, measures to assess the process deployment in projects and identify opportunities for improvement include Compliance/Health Checks, Metric Analysis Reports etc should be defined.

Examples of measures include:

SQA support effort variance

- Frequency of Metric Analysis Reports
- Number of Process Compliance Checks/Health Checks
- Number of early warnings/escalations made to Senior management
- Number of overdue/impending deliverables tracked and corrective actions taken

At this level, the status of SQA work products and delivery of SQA services are visible to the management at defined gates and predefined intervals.

### 6.3 Maturity Level 3

Level 3 establishes the need to have standard processes defined for SQA services, the adherence of which needs to be checked at periodic intervals. The defined process identifies the purpose, roles and responsibilities and Entry Task Validation Exit criteria for SQA services from the project start up throughout the lifecycle. It also facilitates coordination and collaboration with relevant stakeholders from Verticals/Departments.

Based on best practices and feedback from relevant stakeholders, opportunities for improvement are identified on regular basis. Feedback on SQA services may be collected via different forums, for example, senior management meetings, process improvement tools etc.

The extent of customization of SQA services is documented in Tailoring Guideline. Based on effort involved, complexity of project, involvement of vendor and familiarity with process implementation, possible dimensions of tailoring should be defined in Tailoring Guidelines. Project level tailoring is recorded in the SQA plan which is prepared by individual SQA.

Process specific training needs to be imparted to the SQAs prior to their engagement with the projects in facilitating process deployment, so that they can perform their role effectively and efficiently. Apart from process related training, it is also important to identify the need for any tool specific training which would help SQA in the day-to-day engagement with projects.

At this level, Knowledge Management Repository for SQA services may be established based on learning and best practices of SQAs. Challenges faced by SQA and corrective actions taken are maintained in this repository and made available to SQAG as reference towards improvement of their services. This would in-turn serve as an input to SQA plan. Checklists to identify potential issues in projects may be created to help SQAs raise primitive warnings in projects, so that timely actions may be taken. Additionally, risk management database comprising of risks with respect to SQA services will be used an input during preparation of SQA plan.

In cases, where SQAG team comprises of SQAs spread across diversified locations or business units, it is important to organize knowledge sharing sessions/discussions to apprise the members of the team at periodic intervals on the current status.

### 6.4 Maturity Level 4

Level 4 focuses on automating the process of delivering SQA service. This includes usage of tool for the following:

- Tracking of process deployment status and deliverables
- Collection of metrics data and metric analysis
- Scheduling process compliance checks, sharing of relevant documents, preparation of compliance review reports, tracking of non-conformances
- Identify and raise potential issues in collaboration with Project Managers which are automatically escalated to the Senior management for attention depending on the impact

Automation of the aforesaid activities will result in reduction of manual consolidation activities and increase efficiency.

For selected SQA activities, detailed measures of performance are collected over a period of time and statistically analyzed. In addition to automating the process of delivering SQA service, SQA service related metrics are consolidated at an organization level and based on the historical data, metrics are base lined. The baseline thus established will serve as an input for SQA effort estimation in future. Special causes of process variation are identified and where appropriate sources of deviation are identified and corrective actions taken.

### 6.5 Maturity Level 5

Level 5 focuses on reducing the cost spent on deploying the organizations' set of standard processes with minimum SQA facilitation. At this level, Project Managers are expected to be empowered to play role of SQA as well and take accountability of process implementation. The type of process implementation at this level would be referred to as the 'Practitioner Managed Process Implementation' which would ensure institutionalization of processes at a faster rate together with reduction in cost of deployment of processes. It also enables optimal utilization of SQA capacity as SQAs can focus on newly initiated or most critical projects. SQAs can also be further re-deployed to work on process improvement initiatives at the organizational level.

## 7. EXPERIENCE SHARING:

SQA Service Maturity model outlined above was proposed to the customer to justify that mass deployment of processes using 'Practitioner Managed Process Implementation' will be successful only when the organization attains maturity level 5 with respect to SQA Services Maturity Model. The customer was convinced that mass deployment would not be effective with the existing level of process maturity. An implementation plan for implementing SQA service maturity model was prepared along with milestones for attaining target maturity levels.

## 8. CONCLUSION:

This model helps develop competency within the organization to run the process deployment without SQA support byequipping practitioners to take ownership of process deployment. At level 5, lack of independent and objective insight could result in organizational objectives with respect to process improvement being compromised and inconsistent deployment of processes.

It is recommended that the SQA Services Maturity Model be used as a collection of best practices by organizations setting up a Software Quality Assurance Group (SQAG) to drive process implementation in a phased manner. The defined model will provide a framework to empower organizations to choose the appropriate level of SQA services for implementation and institutionalization of defined processes. While this model can serve as a reference model to begin with, practices at Level 4 and Level 5 would also help organizations to reduce cost of SQA activities.

## REFERENCE

[1]   CMMI Product team - CMMI® for Development, by Software Engineering Institute, Carnegie-Mellon University, USA

[2]    Heather Kreger, Tony Carroto (2009) IBM Advantage for Service Maturity Model Standards, http://www.ibm.com/developerworks/webservices/library/ws-OSIMM/index.html

[3]    APMG (2008) - ITIL Service Management Practices: V3 Qualifications Scheme

**Author**

Shankar. G is a Senior Technical Project Leader with Central Research & Development Department in Huawei, Bangalore. He has a total industry experience of about 8 years specializing in Process Definition, Quality Assurance and Business Audits. He has been part of consulting engagements focusing on Agile, ISO standards, CMMI, PCMM models, Metric Analysis & Reporting, for several companies including Deutsche Bank, TCS, NXP Semiconductor, Philips Electronics, Sify Technologies, IC Infotech and Raffles Software. He has presented and published several innovative papers in national and international conferences.

**Publications**

1. "Practitioner Managed Process Implementation Model" presented in NPDC09 conference at IIT Madras

2. Presented "Quantifiable Benefits from CMMI implementation" during the best practice session in Software Process Improvement Network 2009, Bangalore

3. Case study on "How the vendor helps the world leading bank in its Journey of Process Improvement" in TCS newsletter.

4. " Stage Gated QMS Architecture" - IEEE Computer Society 3rd National Conference on Information and Software Engineering (NCISE) 2011, Chennai

5. "Stage Gated QMS for Product Organizations"- International Conference on Operational Excellence for Global Competitiveness (ICOEGC) 2011, Bangalore

6. "Three Dimensional Control Charts" – JSE 2012, Bangalore conducted by AIRCC

7. "Role of QA in Agile" – Agile Conference 2010 in Huawei Technologies, Bangalore

8. "SQA Services Maturity Model" presented in JSE 2012, Bangalore conducted by AIRCC

Link : http://shankargurumoorthy.weebly.com/publications.html

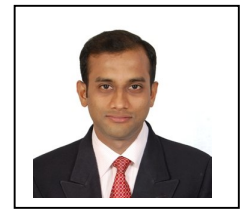